# Genomic Imaging Based on Codongrams and $a^2$grams

E.A. BOUTON[1], H.M. DE OLIVEIRA[1], R.M. CAMPELLO DE SOUZA[1],
N.S. SANTOS-MAGALHÃES[2]

[1]Dep. de Eletrônica e Sistemas, [2]Dep. de Bioquímica, Lab. de Imunologia Keizo-Asami
Universidade Federal de Pernambuco
Caixa postal 7.800 − CDU, 51.711-970, Recife, PE
BRAZIL
{hmo,ricardo,nssm}@ufpe.br   http://www2.ee.ufpe.br/coded/deOliveira.html

**Abstract**: - This paper introduces new tools for genomic signal processing, which can assist for genomic attribute extracting or describing biologically meaningful features embedded in a DNA. The codongrams and $a^2$grams are offered as an alternative to spectrograms and scalograms. Twenty different $a^2$grams are defined for a genome, one for each amino acid (valgram is an $a^2$gram for valine; alagram is an $a^2$gram for alanine and so on). They provide information about the distribution and occurrence of the investigated amino acid. In particular, the metgram can be used to find out potential start position of genes within a genome. This approach can help implementing a new diagnosis test for genetic diseases by providing a type of DNA-medical imaging.

## 1 Introduction

Genes carry information that must be accurately copied and transmitted in live beings. Since the human genome was sequenced [1−4], the genomic signal analysis is attracting wide-ranging interest because of its implication to conceive new diagnosis of diseases, its relevance in the gene therapy and the discovering of new drugs [5,6]. Motivated by the impact of genes for concrete goals −primarily for the pharmaceutical industry − huge efforts have been devoted to exploit DNA sequences.

Protein synthesis involves two steps: transcription and translation. Transcription, which consists of mapping DNA into messenger RNA (*m*-RNA), occurs first; then translation maps the *m*-RNA into a protein, according to the genetic code [7,8]. There are only four different nucleic bases so the code uses a 4-symbol alphabet. DNA sequences are strings of the nucleotides A, T, C, and G. Actually, the DNA information is transcribed into single-strand RNA−the mRNA. In this circumstance, thymine (T) is replaced by the uracil (U). The information is transmitted by a start-stop protocol. The genetic source is entirely characterized by the alphabet $\aleph:=\{U,C,A,G\}$. The input alphabet $\aleph^3$ is the set of 'codons' $\aleph^3:=\{c_1,c_2,c_3 \mid c_i \in \aleph, i=1,2,3\}$. The output alphabet $A$ is the set of amino acids including the nonsense 'codons' (Stop elements): $A:=\{Leu, Pro, Arg, Gln, His, Ser, Phe, Trp, Tyr, Asn, Lys, Ile, Met, Thr, Asp, Cys, Glu, Gly, Ala, Val, Stop\}$. Therefore, the genetic code maps the 64 radix-3 'codons' (5' to 3' DNA) of the DNA characters into one of the 20 possible amino acids (or into a punctuation mark). The genetic code is a mapping $GC: \aleph^3 \to A$ that maps triplets $(c_1,c_2,c_3)$ into one amino acid $A_i$. For instance, $GC(UAC)=Stop$ and $GC(CAC)=His$. New representations for the GC were recently introduced [9]. Let $\|.\|$ denote the cardinality of a set. Evaluating the cardinality of the input and the output alphabets, we have, $\|\aleph^3\|=\|\aleph\|^3=64$ and $\|A\|=21$, thereby showing that GC is an extensively degenerated code.

The aim of this paper is to introduce new tools for genomic signal analysis (GSA) [10]. Specifically, two new representations of genome, namely 'codongrams' and 'a²grams', are defined. The idea behind these spectrum-like diagrams is to perform genome decomposition. The 'codongram' describes the distribution of a 'codon' through the genome. The 'a²gram' for a particular amino acid provides information about the sections of the DNA strand, which potentially leads to the synthesis of such an amino acid. DNA 'codongrams' and 'a²grams' are among powerful visual tools for GSA like spectrograms and scalograms [11,12], which can be applyied when searching for particular nucleotide patterns.

## 2 Codon Representation and Codon Inner Product

Each 'codon' is represented by a triplet $c_1,c_2,c_3$, where $c_i$ are nucleotides, $c_i \in N:=\{A,T,C,G\}$, $i=1,2,3$. Nucleotides can be replaced by binary labels according to the rule ($x \to y$ denotes the operator *replace x by y*): T→[11]; A→[00]; G→[10]; C→[01].

The usefulness of this specific labeling can be corroborated by the following argument. The 'complementary base pairing' property can be interpreted with the aid of binary labels as some sort of parity check. There are several relations between DNA and error-correcting codes [9,13]. The DNA parity can be defined as the modulo 2 sum of all binary coordinates of the nucleotide representations (Table 1). As expected, the only checked parities are A=T and C≡G (biochemistry chemical notation), the main diagonal apart.

Table 1. Parity-check on nucleotides of the DNA.

| parity check | [00]←A | [01]←C | [10]←G | [11]←T |
|---|---|---|---|---|
| [00]←A | even | odd | odd | even |
| [01]←C | odd | even | even | odd |
| [10]←G | odd | even | even | odd |
| [11]←T | even | odd | odd | even |

Labelling for paired bases in a DNA strand gives an error-correcting code as shown in Table 2. Each binary codeword belongs to a constant weight code.

Table 2. Double-strand DNA short section of the icosahedral bacterial virus ΦX174 [14]: base-pairs and their corresponding binary label sequences.

| DNA | Codeword | |
|---|---|---|
| G…C | 10 | 01 |
| A…T | 00 | 11 |
| G…C | 10 | 01 |
| T…A | 11 | 00 |
| T…A | 11 | 00 |
| T…A | 11 | 00 |
| T…A | 11 | 00 |
| A…T | 00 | 11 |
| T…A | 11 | 00 |
| G…C | 10 | 01 |
| G…C | 10 | 01 |
| C…G | 01 | 10 |
| T…A | 11 | 00 |

The complementary nucleotides are defined to match with the hydrogen bonds that appear in the 'complementary base pairing'. They are computed by the operator * in such a way that A*=T, T*=A, C*≡G, and G*≡C.

**Definition 1** (*anticodon*). The 'anticodon' $\underline{c}^* \in N^3$ of a 'codon' $\underline{c} = (c_1, c_2, c_3)$, $c_i \in N$, is defined by $\underline{c}^* := (c_1^*, c_2^*, c_3^*)$. ■

The 'anticodon' corresponds to applying a NOT gate of the corresponding binary labels. For instance, $\underline{c}$=(A,T,C) and $\underline{c}^*$=(T,A,G) are complementary 'codons'. Their binary (0,1)-valued representations are consequently
$\underline{c}$=(0 0 1 1 0 1) and $\underline{c}^*$ =(1 1 0 0 1 0)=NOT($\underline{c}$).

Another interesting new concept is the 'anti-amino acid'. All triplets ('codons') that encode the same amino acid will be referred to as its homophones. Given an amino acid, their anti-amino acid corresponds to the amino acid set generated by the anti-homophones. For example, {GCG, GCA, GCC, GCU} are homophones for alanine. The set of its anti-homophones is {CGC, CGU, CGG, CGA}, which are always translated into arginine. Table 3 shows the anti-amino acids of each amino acid.

Table 3. Anti-amino acids.

| Complementary amino acid |
|---|
| Ala* = Arg |
| Arg* = (Ala, Ser) |
| Asn* = Leu |
| Asp* = Leu |
| Cys* = Thr |
| Gln* = Val |
| Glu* = Leu |
| Gly* = Pro |
| His* = Val |
| Ile* = (Stop, Tyr) |
| Leu* = (Asn, Glu, Asp) |
| Lys* = Phe |
| Met* = Tyr |
| Phe* = Lys |
| Pro* = Gly |
| Ser* = (Ser, Arg) |
| Stop* = (Ile, Thr) |
| Thr* = (Cys, Stop, Trp) |
| Trp* = Thr |
| Tyr* = (Met, Ile) |
| Val*=(His, Gln) |

A more suitable binary labelling of nucleotides to define an inner product should be
T→ [1 1]; A → [-1 -1]; G →[1 -1]; C →[-1 1].

**Definition 2** (*'codon' inner product*). An inner product between two 'codons' $\underline{c}_1$ and $\underline{c}_2$ can be induced by the usual inner product between their corresponding binary labels, that is,
$<\underline{c}_1, \underline{c}_2> := <(c_{1,1}, c_{1,2}, c_{1,3}), (c_{2,1}, c_{2,2}, c_{2,3})>$. ■
For instance, the inner product between AGT and TTG, denoted by < A G T, T T G>, is given by:
<AGT,TTG>=<(-1 -1 1 -1 1 1),(1 1 1 1 1 -1)>=–2.
Some direct properties of the 'codon inner product' follow from this definition:

**Properties**
**P1**. The 'codon inner product' is commutative, i.e.,
$<\underline{c}_1, \underline{c}_2> = <\underline{c}_2, \underline{c}_1>$, $(\forall \underline{c}_1, \underline{c}_2 \in N^3)$.
**P2**. The inner product between two 'codons' is one of the integers of the set I:={0, ±2, ±4, ±6}.

**P3**. The inner product between a 'codon' and itself achieves the maximum value
$$<\underline{c}, \underline{c}> = 6 \quad (\forall \underline{c} \in N^3).$$
**P4**. The product of a 'codon' by its complementary 'codon' reaches the minimum value,
$$<\underline{c}, \underline{c}^*> = -6 \quad (\forall \underline{c} \in N^3).$$
**P5**. If none of the corresponding nucleotides of two 'codons' are identical or complementary then they are orthogonal, i.e., $<\underline{c}_1, \underline{c}_2> = 0$.

The 'codon inner product' can also be computed by means of the Hamming distance between their nucleotides: $<\underline{c}_1, \underline{c}_2> = 2 \cdot (D_H(\underline{c}_1, \underline{c}_2^*) - D_H(\underline{c}_1, \underline{c}_2))$ where $D_H$ is the Hamming distance.

**P6**. Two 'codons' $\underline{c}_1$ and $\underline{c}_2$ are orthogonal if and only if they are half-way between being identical or complementary, that is, $D_H(\underline{c}_1, \underline{c}_2) = D_H(\underline{c}_1, \underline{c}_2^*)$.

## 3 Genomic Analysis Based on the Product of DNA Sequences

The number of base pairs of a DNA in its haploid genome is usually referred to as the *C*-value of the genome. This concept is especially useful to discuss the *C*-value paradox [14, p.1133]. Typically, the *C*-value does not divide 3, so the genome does not have, necessarily, an integer number of 'codons'.

A few operations are introduced below in order to handle genomic sequences.

**Operation 1** (*genomic padding*). A single-strand DNA is first converted into a DNA vector $g$ = (the genomic sequence) by appending -*C* (*mod* 3) zeroes to the original sequence. ∎

Whether zeroes must or not be appended depends on the original length of the genome. This operation is necessary to guarantee that the number of components (nucleotides or stuffing) of the vector $g$ is divisible by 3. The *C*-value of the padded-sequence becomes $\|g\| = 3\lceil C/3 \rceil$, where $\lceil \ \rceil$ denote the classical ceiling function.

The genomic vector $g$ is given by
$$\underline{g} = (c_1^1 c_2^1 c_3^1 \ c_1^2 c_2^2 c_3^2 \dots c_1^{\lceil C/3 \rceil} c_2^{\lceil C/3 \rceil} c_3^{\lceil C/3 \rceil}),$$
that is, $\underline{g} = (\underline{c}^1 \ \underline{c}^2 \ \underline{c}^3 \ \underline{c}^4 \dots \underline{c}^{\lceil C/3 \rceil})$.

The components of $g$ are therefore 'codons' or 'pseudo-codons', the latter including at least a stuffing nucleotide.

**Example 1**
The genomic padding process is illustrated below by analysing a hypothetically short DNA double-strand
  5' A G T C G T C C A A G T C 3'
  3' T C A G C A G G T T C A G 5'
This genome has a *C*-value 13 so 2 null-components should be appended (stuffing). Therefore, $g$ = (A G T C G T C C A A G T C 0 0).

If the analysed genome is cyclic as it often occurs in most viruses, the stuffing components must be done by repeating the start portion of the DNA string instead of appending zeroes. In the above example, the genomic sequence should be
  $g$ = (A G T C G T C C A A G T C A G). □

Another genomic operation is defined in order to deal with the different reading frames of the DNA sequence [8,14].

**Operation 2** (*reading frames*). Three sequences at different reading frames (*rf*) can be generated from a given genome sequence $g$, namely $\underline{\dot{g}}$, $\underline{\ddot{g}}$, $\underline{\dddot{g}}$. They are compatible with cyclic shifts of the original sequence. If $D$ denotes the cyclic shift operator, then
$$\underline{\dot{g}} = \underline{g}, \quad \underline{\ddot{g}} = D^2(\underline{g}), \quad \underline{\dddot{g}} = D(\underline{g}). \ \blacksquare$$

The number of dots corresponds to the reading frame of the sequence $g$.

**Example 1** (*revisited*)
The short genome presented in example 1 possesses the following reading frame sequences:
$\underline{\dot{g}}$ = (A G T : C G T : C C A : A G T : C A G).
$\underline{\ddot{g}}$ = (A G A : G T C : G T C : C A A : G T C).
$\underline{\dddot{g}}$ = (G A G : T C G : T C C : A A G : T C A). □

The inner product of DNA strings of same length is defined by means of the 'codon inner product'.

**Definition 3** (*Inner product of DNA sequences*). Given two DNA sequences of length $3\lceil C/3 \rceil$, say,
$$\underline{g}_1 = (\underline{c}^{1,1} \ \underline{c}^{1,2} \ \underline{c}^{1,3} \dots \underline{c}^{1,\lceil C/3 \rceil}),$$
$$\underline{g}_2 = (\underline{c}^{2,1} \ \underline{c}^{2,2} \ \underline{c}^{2,3} \dots \underline{c}^{2,\lceil C/3 \rceil}),$$
the inner product $\underline{g}_1 \bullet \underline{g}_2$ is defined by
$$\underline{g}_1 \bullet \underline{g}_2 := \sum_{j=1}^{\lceil C/3 \rceil} <\underline{c}^{1,j}, \underline{c}^{2,j}>. \ \blacksquare$$

Here, $\underline{c}^{i,j} = (c_1^{i,j}, c_2^{i,j}, c_3^{i,j}) \in N^3$ are 'codons' or 'pseudo-codons', for $i=1,2; j=1,2,\dots,\lceil C/3 \rceil$.

**Example 2**
The inner product between the DNA-sequences A T C T G C C G A and A C G G G T A T T is
$\underline{g}_1 \bullet \underline{g}_2$ = <ATC,ACG>+<TGC,GGT>+<CGA,ATT>
If $\underline{g}_1 \bullet \underline{g}_2 = 0$ then the genomes $\underline{g}_1$ and $\underline{g}_2$ are said to be orthogonal DNA, as usual. □

Besides the inner product between DNA sequences, a vector product can also be defined.

**Definition 4** (*vector product of DNA sequences*). Given two DNA sequences $\underline{g}_1$ and $\underline{g}_2$ of length $3\lceil C/3 \rceil$ (the padding operation could be required),

$\underline{g}_1 = (\underline{c}^{1,1}\ \underline{c}^{1,2}\ \underline{c}^{1,3}\ ...\ \underline{c}^{1,\lceil C/3 \rceil})$

$\underline{g}_2 = (\underline{c}^{2,1}\ \underline{c}^{2,2}\ \underline{c}^{2,3}\ ...\ \underline{c}^{2,\lceil C/3 \rceil})$,

the vector product $\underline{g}_1 \otimes \underline{g}_2$ is a vector with the same length of both genomes given by $\left( <\underline{c}^{1,1}, \underline{c}^{2,1}>, <\underline{c}^{1,2}, \underline{c}^{2,2}>, ..., <\underline{c}^{1,\lceil C/3 \rceil}, \underline{c}^{2,\lceil C/3 \rceil}> \right)$. ∎

The vector product $\underline{g}_1 \otimes \underline{g}_2$ has thus $\lceil C/3 \rceil$ I-valued coordinates.

## 4 Codon-Finder Sequences and Amino Acid Localisers

Suppose that all the occurrences of a particular 'codon' $\underline{c} = c_1 c_2 c_3$ must be found within a given genomic sequence $g$.

**Definition 5** (*codon-finder sequence*). The '$\underline{c}$-codon-finder' sequence $\underline{F}(\underline{c})$ is a sequence of same length as $g$ in which $c_1 c_2 c_3$ is repeated until it achieves the length of $g$. ∎

**Example 3** (AGT-*finder sequence*).
Let $g$ be the genomic sequence of the example 1.
$g$ = (A G T C G T C C A A G T C 0 0), so the AGT-finder sequence at the reading frame 1 is the sequence
$\underline{F}$(AGT) :=(A G T A G T A G T A G T A G T).

The 'codon-finder sequence' may also take into account a specific reading frame in which the 'codon' $\underline{c} = c_1 c_2 c_3$ must be found. It suffices to include suitable pseudo-nucleotides (stuffing nucleotides) as shown below.
$\underline{F}^{\{rf=1\}}$(AGT) :=(A G T A G T A G T A G T A G T),
$\underline{F}^{\{rf=2\}}$(AGT) :=(G T A G T A G T A G T A G T A),
$\underline{F}^{\{rf=3\}}$(AGT) :=(T A G T A G T A G T A G T A G).
The value of the parameter *rf* selects the reading frame. ∎

The 'codon-finder sequence' is used to help identifying the positions in the padded-sequence where the 'codon' takes place.

**Definition 6** (*codon-localiser*). Let $\underline{c}$ be the 'codon' to be localised. The '$\underline{c}$-localiser' at the *rf*-th reading frame is a vector defined by the vector product between the genomic sequence and the 'codon-finder sequence' at the reading frame *rf*, i.e.,

$\underline{L}^{\{rf\}}(\underline{F}(\underline{c}),g) := \underline{F}^{\{rf\}}(\underline{c}) \otimes \underline{g}$. ∎

The 'codon-localiser' will be denoted by $\underline{L}_{\underline{c}}^{\{rf\}}$.

**Example 4**
The AGT-localisers for the genome given in the example 1 are

$\underline{L}_{AGT}^{\{1\}} = \underline{L}^{\{1\}}(\underline{F}(AGT), \underline{g}) := (6\ 4\ -4\ 6\ 0)$,
$\underline{L}_{AGT}^{\{2\}} = \underline{L}^{\{2\}}(\underline{F}(AGT), \underline{g}) := (-2\ -4\ 0\ -2\ -2)$,
$\underline{L}_{AGT}^{\{3\}} = \underline{L}^{\{3\}}(\underline{F}(AGT), \underline{g}) := (-2\ 0\ 0\ -2\ 0)$.

For details, see example 3. For instance $\underline{L}^{\{1\}}$ is given by (AGT AGT AGT AGT AGT)⊗(AGT CGT CCA AGT C00). ∎

**Operation 3** (*combining 'codon–localisers'*).
Let $\underline{L}_1 = (L^{1,1}, L^{1,2}, ..., L^{1,\lceil C/3 \rceil})$ and $\underline{L}_2 = (L^{2,1}, L^{2,2}, ..., L^{2,\lceil C/3 \rceil})$, $L^{i,j} \in I$, $i$=1,2, $j$=1,2,..., $\lceil C/3 \rceil$ be two 'codon-localisers'.

They can be combined according to the following rule: $\underline{L} := \underline{L}_1 \oplus \underline{L}_2$ is a new localising vector of same length as $\underline{L}_i$, whose coordinates are the maximum value between the corresponding coordinates of $L_1$ and $L_2$, that is, $\underline{L} := (L_1, L_2, ..., L_{\lceil C/3 \rceil})$, where $L_j := MAX(L^{1,j}, L^{2,j})$. ∎

The AGT-localiser at different reading frames as shown in example 4 can be combined into a single AGT-localiser:

$\underline{L}_{AGT} := \underline{L}_{AGT}^{\{1\}} \oplus \underline{L}_{AGT}^{\{2\}} \oplus \underline{L}_{AGT}^{\{3\}} = (6\ 4\ 0\ 6\ 0)$.

This operation preserves the positions where the 'codon' was found regardless the reading frame. Three localisers are then collapsed into a single one. The 'codongram' is a visual representation of the 'codon-localiser'. The one-dimensional vector $\underline{L}$ is firstly rearranged into a two-dimensional array. The number of rows and columns is given by $\sim \sqrt{\lceil C/3 \rceil}$. The 'codongram' is derived by associating a colour map with the array corresponding to the localiser $\underline{L}$. The combining operation of 'codon-localisers' is mainly useful to deal with amino acids. As previously discussed, the genetic code is highly degenerated and several 'codons' yield the same amino acid.

**Definition 7** (*amino acid localisers*). The '$a^2$-localiser' for an amino acid $A_i$ at a reading frame *rf* is defined by

$$\underline{L}_{A_i}^{\{rf\}} := \sum_{GC(c_1 c_1 c_3) = A_i}^{\circ} \underline{L}^{\{rf\}}(F(c_1, c_2, c_3), \underline{g}).$$

The sum $\sum^{\circ}$ denotes a summation using the special "addition" defined in operation 3.

For instance, $\underline{L}_{STOP}^{\{1\}} = \underline{L}_{TAA}^{\{1\}} \oplus \underline{L}_{TGA}^{\{1\}} \oplus \underline{L}_{TAG}^{\{1\}}$.

Localisers of an amino acid at different reading frames can thus be combined. The 'amino acid full-localiser' is thereby the mixture of all 'codon-localisers' that are homophones of such an amino acid, not considering the reading frame,

$$\underline{L}_{A_i} := \sum_{rf=1}^{3} \circ \sum_{GC(c_1c_2c_3)=A_i} \circ \underline{L}^{\{rf\}}_{c_1c_2c_3}.\blacksquare$$

Given a newly sequenced genome, suppose that the start positions of genes are to be located. We are looking for incidence of the 'codon' that is coded into the methionine. Since GC(AUG)=*Met*,

$$\underline{L}_{MET} = \sum_{rf=1}^{3} \circ \underline{L}^{\{rf\}}_{ATG}$$, where $\underline{L}^{\{rf\}}_{ATG} = \underline{L}^{\{rf\}}(\underline{F}(ATG), g)$, for *rf*=1,2,3. The STOP-localiser, for instance, is

$$\underline{L}_{STOP} = \underline{L}^{\{1\}}_{STOP} \oplus \underline{L}^{\{2\}}_{STOP} \oplus \underline{L}^{\{3\}}_{STOP},$$

where $\underline{L}^{\{rf\}}_{STOP} = \underline{L}^{\{rf\}}_{TAA} \oplus \underline{L}^{\{rf\}}_{TGA} \oplus \underline{L}^{\{rf\}}_{TAG}$.

The 'a²-localiser' can be plotted as an 'a²gram' in the same way as done for 'codongrams'. The tag of the 'a²gram' is specified according to the amino acid under investigation: The valgram is an 'a²gram' for valine; the alagram is an 'a²gram' for alanine, etc. A few a²gram for the phague ΦX174 [15] are shown in Fig. 1.

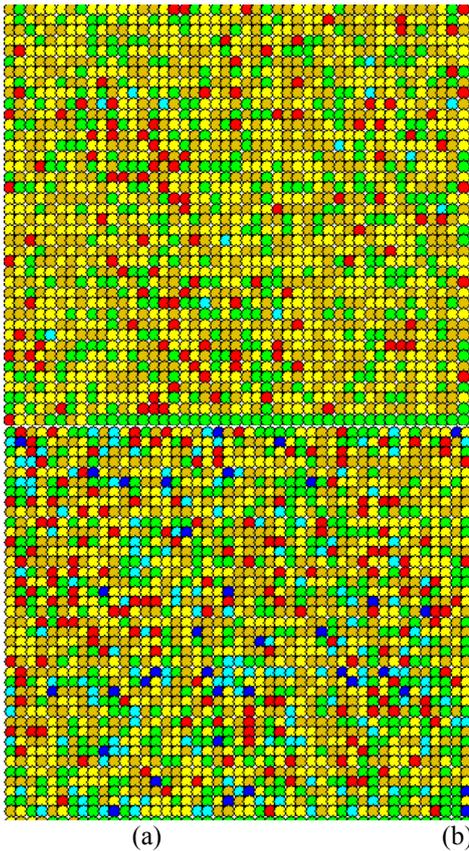

(a)  (b)

Figure 1. a²gram for the ΦX174 virus: a) metgram; b) phegram. "Hot" points correspond to maximum value. Colour map:{-6(magenta), -4(blue), -2(cyan), 0(green), 2(yellow), 4(orange), and 6(red)}.

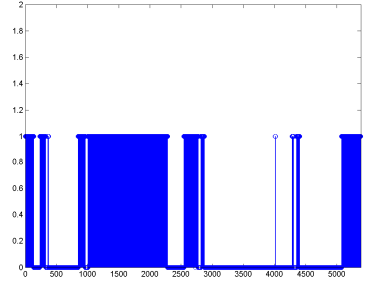

Figure 2. Start-Stop-Mask for ΦX174 at rf=2. Abscissa 1-5386 indicates the nucleotide position in the RNA. The binary mask is 1 if the position is between a start and a stop, zero otherwise.

Setting a particular reading frame, a start-stop-mask is generated from the localizer sequences $\underline{L}_{ATG}^{\{rf\}}$ and $\underline{L}_{Stop}^{\{rf\}}$. It is a binary sequence starting with zero, which is set to 1 after each position where $\underline{L}_{ATG}^{\{rf\}} = 6$, and is forced to 0 after reaching a position where $\underline{L}_{Stop}^{\{rf\}} = 6$. The mask turns black all the positions of 'noncoding codons'. Figure 2 shows start-stop mask derived for the coliphage ΦX174. The mask can be applied to any a²gram. Figure 3 shows the Metgram applied to the known RNA-cyclic bacterial virus ΦX174, which has only ten genes [14], after submitted to the start-stop-mask.

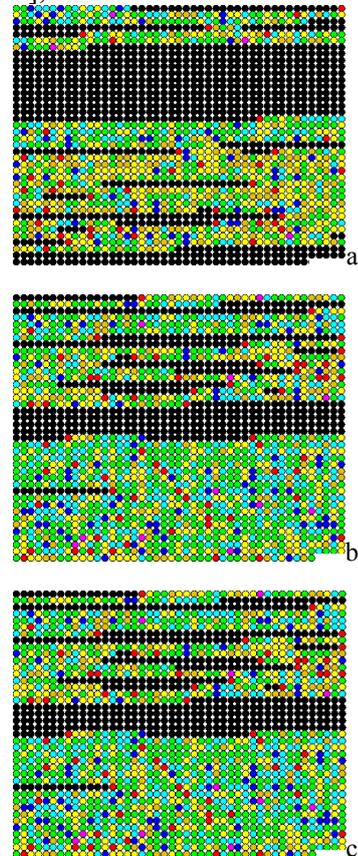

Figure 3. Metgram for the phage ΦX174 at different reading frames (rf). Genes start with a "hot" point and finish with a "black" point. a) rf=2; b) rf=1; c) rf=3.

**Definition 8**. If $S_1$ and $S_2$ are start-stop strings then $S_1$ is said to be covered by $S_2$, *if and only if* $S_1$ can be found in $S_2$. ∎

Start-stop strings control the synthesis of proteins. However, besides start-stop portions corresponding to the genes of the phage (Table 4), there exists many short start-stop regions that do not specify genes.

## 5   Conclusions

The evaluation of a patient´s health is, as a rule, involved in the higher-level, location-based interpretation on DNA sequences. Transform-based DNA imaging has a computational complexity by far much greater than a$^2$grams, since the latter only requires integer additions. Much of GSA approaches for genomic feature extraction and functional cataloguing have been focused on oligonucleotide patterns in the linear primary sequences of genomes [5,6,10,16]. Analysing a specific chromosome, explicit patterns will emerge due to a genetic disease. The new tools, 'codongrams' and 'a$^2$grams', are DNA-medical imaging, which can be applied for the human DNA. Several neurological diseases are associated with trinucleotide repeat expansion (fragile X syndrome, myotonic dystrophy, Huntington's disease etc.) [14]. For instance, the Huntington's disease, a devasting neurodegenerative disorder, typically appears after an age of ~40 years. It is characterized by a polymorphic (CAG)$_n$ repeat more than 40 times in the chromosome 4. In contrast, this nucleotide sequence is only repeated about 6 times for people not affected by this disorder. The 'condon-finder sequence' $\underline{F}_{CAG}$ could be used for analysing the DNA-content of the chromosome 4. As a result, a codongram-based diagnosis test can be implemented.

*Acknowledgements*: this study was partially supported by the Brazilian National Council for Scientific and Technological Development (CNPq) under research grants N.306180 (HMO) and N.306049 (NSSM).

## Appendix

Table 4. Potential genes of the coliphage ΦX174 according to the 'gene-localiser' operator. Only single-strand sequences between a start and a stop that yield more than 50 amino acids are shown. The genes A and B have two different reading frames. This happens because 5386 does not divide 3.

| Position | Protein length (aa) | Gene | Reading frame |
|---|---|---|---|
| 51 - 219 | 56 | K | 3 |
| 390 - 846 | 152 | D | 3 |
| 1038 - 1194 | 52 | None ⊂ F | 3 |
| 1599 - 1773 | 58 | None ⊂ F | 3 |
| 1998 - 2232 | 78 | None ⊂ F | 3 |
| 2931 - 3195 | 328 | H | 3 |
| 3981 (rf3) – 135 (rf2) | 455 | A | 3, 2 |
| 849 - 963 | 38 | J | 2 |
| 1002 - 2283 | 427 | F | 2 |
| 2544 - 2730 | 62 | None ⊂ G | 2 |
| 5076 (rf2) – 51 (rf1) | 120 | B | 2, 1 |
| 132 - 390 | 86 | C | 1 |
| 567 - 840 | 91 | E | 1 |
| 2394 - 2919 | 175 | G | 1 |
| 3075 - 3681 | 202 | Unknown | 1 |
| 3741 - 3927 | 62 | Unknown | 1 |
| 3945 - 4260 | 105 | Unknown | 1 |
| 4620 - 4854 | 78 | none ⊂ A | 1 |
| 4881 - 5061 | 60 | none ⊂ A | 1 |